\documentstyle[prl,aps,amsmath,amssymb,epsfig]{revtex}

\newcommand{\etal}{\emph{et al}.}

\title{Excitations of the field-induced soliton lattice in CuGeO$_3$\\}
\author{M. Enderle$^{1,2}$, H. M. R\o{}nnow$^{2,3}$, D. F. McMorrow$^2$,
L.-P. Regnault$^3$,\\ G. Dhalenne$^4$, A. Revcholevschi$^4$, P.
Vorderwisch$^5$, H. Schneider$^{6}$, P.~Smeibidl$^5$, and
M.~Mei\ss ner$^5$}
\address{$^1$Technische Physik, Universit\"at des Saarlandes,
66123 Saarbr\"ucken,
Germany\\ $^2$Condensed Matter Physics and Chemistry Department,
Ris\o{} National Laboratory, 4000 Roskilde, Denmark\\
$^3$DRFMC, CENG, CEA, 38054 Grenoble, France\\ $^4$Laboratoire de
Chimie des Solides, Universit\'{e} de Paris Sud, Orsay, France\\
$^5$BENSC, 14109 Berlin, Germany\\$^6$ Technische Universit\"at
M\"unchen, Germany}

\date{\today}

\begin{document}
\twocolumn[\hsize\textwidth\columnwidth\hsize\csname
@twocolumnfalse\endcsname
\maketitle
\begin{abstract}
Here we report the first inelastic neutron scattering study of the
magnetic excitations in the incommensurate high-field phase of a
spin-Peierls material. The results on CuGeO$_3$ provide direct
evidence for a finite excitation gap, two sharp magnetic
excitation branches and a very low-lying excitation
which is identified as a phason mode, the Goldstone mode of the
incommensurate soliton lattice.

\end{abstract}
 ] A
one-dimensional (1D) spin $\frac12$ Heisenberg antiferromagnet is
unstable with respect to dimerization as well as long-range
antiferromagnetic order. Coupled to a three-dimensional (3D)
phonon field, it can undergo a second-order phase transition at a
finite temperature into a dimerized (spin-Peierls) ground state
with total spin $\vec{S}_{tot}=\sum_i \vec{S}_i =0$ (D-phase).
The dimerization is supported by antiferromagnetic
next-nearest-neighbor exchange in the 1D chain direction.
Antiferromagnetic (AF) interchain couplings prefer a
N\'{e}el-type ground state and compete with dimerization.

In the D-phase of a spin-Peierls material, the lowest magnetic
excitation is an isolated triplet branch ($S_{tot}=1$). This
excitation is a bound pair of domain walls with respect to the
dimer order parameter. The two domain walls, or solitons, are
created by breaking and delocalizing one dimer bond. The binding
energy depends on the magnetic and elastic interchain
interactions. The energy of the triplet is finite over the entire
Brillouin zone and achieves its minima at wave vectors $k=0$ and
$k=\frac{\pi}c$ (the AF zone center, $c$ being the average
distance between two spins in the chain direction.) In a magnetic
field, the $S_{tot}=1$ branch becomes Zeeman split, and a
magnetized incommensurate (IC) phase is entered above a critical
field $H_c$, which corresponds approximately to the field where
the lowest mode should soften to zero energy. In the IC-phase,
the magnetization is generated by introducing more and more domain
walls into the dimerized ground state. The unpaired spin at the
center of a domain wall aligns parallel to the magnetic field
(assumed to be $\parallel$ to $z$). The transverse parts of the
spin Hamiltonian, $S^+_i S^-_j$, delocalize the solitons. This
results in an equal spacing of solitons, and thus leads to an IC
superlattice of distortive and magnetic solitons. Their finite
width implies a staggered magnetic polarization close to the
soliton center.

For a spin-Peierls system, the magnetic excitation spectrum in the
IC-phase is expected to have finite excitation gaps, because at
each given magnetic field the lattice, and hence the intrachain
exchange, adapts to the magnetization and the number of solitons.
Two excitation branches with polarization perpendicular to $H$
are predicted, $\Delta_{\pm}$, corresponding to an increase or
decrease of the total spin by 1, with minima at $k=m
\frac{2\pi}{c}$ and $k=\frac{\pi}{c}$, where $m$ is the
magnetization per spin in units of $\mu_B$ \cite{uhr98a}. A
longitudinal branch $\Delta_0$ should also be present with minima
at $k=0$ and $k=\frac{2\pi}{c}(\frac{1}{2}-m)$, and with
$\Delta_0^{min}\gtrsim\frac{1}{2}(\Delta_+^{min}+\Delta_-^{min})$
\cite{yu99}. At $k=\frac{\pi}{c}$ this branch should have an
energy $g\mu_B H$ \cite{uhr98a}. The precise value and field
dependences of the modes $\Delta_{\pm}$ and $\Delta_0$ depend on
the details of the assumptions used in the
models\cite{yu99,uhr99b,sch98d,loo96a}.

CuGeO$_3$ is the first spin-Peierls compound where large single
crystals are available \cite{has93a}, thus allowing the complete
phonon and magnetic excitation spectra to be measured  using
neutron scattering techniques. The IC phase is also accessible,
since magnetic fields up to 14.5~T have become available for
neutron experiments, well in excess of the critical field in
CuGeO$_3$ of $\mu_0H_c\approx 12.5$~T. Our recent neutron
scattering study of the magnetic soliton lattice in the IC-phase
\cite{roe00} reveals a static magnetic modulation with the same
period as the distortive modulation, and in particular the
superlattice Bragg peaks at $(\frac12,1,\frac12\pm\delta k_{sp})$
were found to be almost entirely magnetic in origin. Outside the
critical field region, the magnetic soliton width, and the
amplitudes of the staggered and uniform magnetic modulation are
in reasonable agreement with the predictions of field theory.

In this letter we report on the magnetic excitation spectrum in
the IC phase, measured both  close to the superlattice peak
$(\frac12,1,\frac12\pm\delta k_{sp})$, and also at the AF zone
center $(0,\frac12,1)$, where the dispersion of the magnetic
excitations in the D-phase has its minimum. The experiments were
performed at the cold triple axis spectrometer FLEX at BENSC,
Berlin, using the vertical cryomagnet VM-1. Single crystals of
CuGeO$_3$ of 0.34~cm$^3$ and 0.49~cm$^3$ were aligned in the
$(0,k,\ell)$ and $(h,2h,\ell)$ scattering planes, respectively.
The collimation was guide-60'-60'-60' with either $k_f$=1.3~\AA\
fixed and a nitrogen-cooled Be-filter in $k_i$, or
$k_f=1.54$~\AA\ and the Be-filter in $k_f$. The critical fields
at $(0,1,\frac12)$ and $(\frac12,1,\frac12)$ differ (12.58~T and
12.31~T) because of the different field directions and hence
$g$-factors (2.152 and 2.199) in the respective scattering
geometry.

\begin{figure}
\begin{center}
\epsfig{file=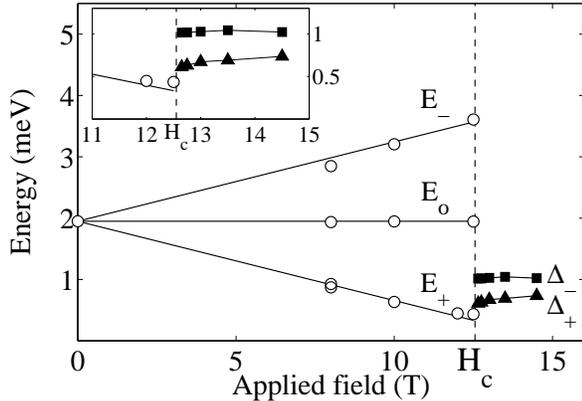,width=0.9\columnwidth,angle=0}
\caption[]{Field dependence of the triplet excitation in CuGeO$_3$
at $(0,1,\frac12)$, 2~K. Solid lines: triplet splitting $g_a\mu_B
H S^a_{tot}$, $g_a=2.152$. The critical field is $\mu_0
H_c=12.58$~T. The inset shows the high-field region in more
detail.} \label{fig1}
\end{center}
\end{figure}

Figure \ref{fig1} shows the magnetic field dependence of the
excitations at the AF zone center $(0,1,\frac12)$. With
increasing field, the lowest triplet branch decreases to about
0.4~meV but never becomes completely soft. At the critical field,
two new sharp excitations appear at $\sim$1~meV and $\sim$0.5~meV.
The 1~meV mode shows barely any field dependence between 12.5~T
and 14.5~T, while the energy of the lower mode increases towards
the higher mode. The excitations at $(\frac12,1,\frac12)$, close
to the magnetic Bragg peaks at $(\frac12,1,\frac12\pm\delta
k_{sp})$, display the same field dependence as those at
$(0,1,\frac12)$, apart from an overall shift towards higher
energies (1.2~meV and 1.6~meV) which is entirely accounted for by
the ferromagnetic interchain interaction along $a^{\ast}$.

In Fig.\ \ref{fig2} we show high-resolution scans of the
low-energy region at a wave vector of $(\frac12,1,\frac12)$ for
selected magnetic fields. At 12~T, the widths of the middle and
the lowest triplet excitation are resolution limited. Both broaden
significantly at the critical field (about 12.3~T) and then
disappear. In a small field region, the new sharp IC-phase
excitations at 1.2~meV and 1.6~meV coexist with the broadened
D-phase modes (see scan at 12.5~T), confirming the first-order
character of the transition which has also been observed with
other methods. No evidence was found for a third excitation at an
energy of $g\mu_BH$ at any of the fields investigated.

The spectral weight of the two new modes in the IC-phase is
concentrated in a very small region extending from the
commensurate position $\ell=\frac12$ to the incommensurate one at
$\ell=\frac12\pm\delta k_{sp}$. The upper part of Fig.\
\ref{fig3} shows the dispersion of the lowest two excitations at
$(\frac12,1,\frac12)$ along the chain direction $c^{\ast}$, at
$T=2$~K and $B=14.5$~T, where $\delta k_{sp}=0.015$. The
dispersion (solid lines) appears significantly flatter than the
dispersion calculated from the zero field spin-wave velocity
(dashed lines). Neither at $(0,1,\frac12)$ nor at
$(\frac12,1,\frac12)$ do the dispersions of the two sharp modes
exhibit minima at the IC positions $\ell=\frac12\pm\delta k_{sp}$.
\begin{figure}[htb!]
\begin{center}
\epsfig{file=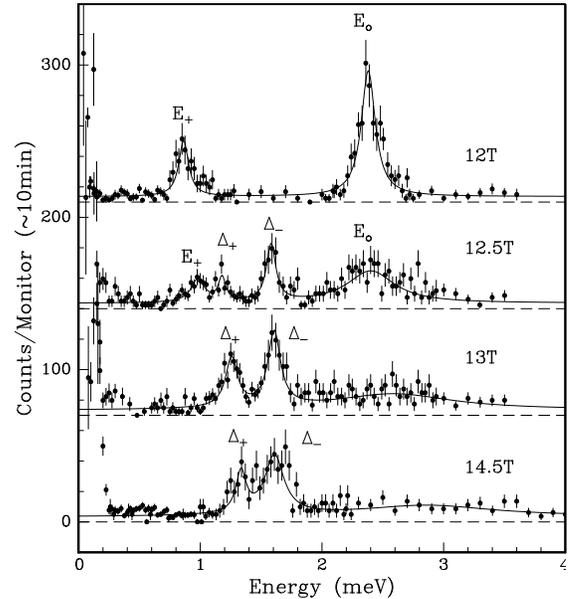,width=0.9\columnwidth,angle=180}
\caption[]{Energy scans at $(\frac12,1,\frac12)$, T=2~K. Solid
lines: fits to Lorentzian-type profiles including the Bose
factor. The critical field is $\mu_0 H_c=12.31$~T.} \label{fig2}
\end{center}
\end{figure}

\begin{figure}[htb!]
\begin{center}
\epsfig{file=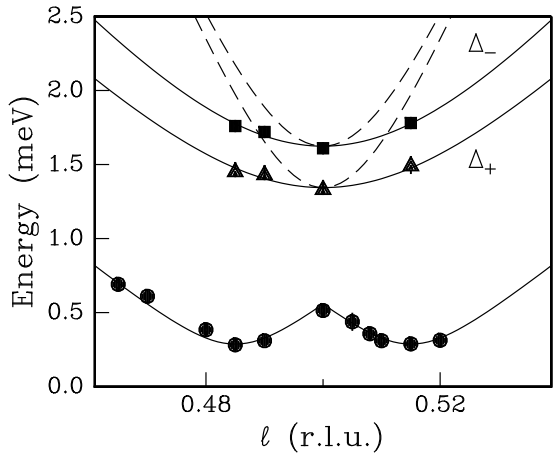,width=0.9\columnwidth,angle=180}  %
\caption{$(\frac12,1,\ell)$-dispersion of the lowest magnetic
excitations in CuGeO$_3$ at 14.5~T, 2~K. Solid: dispersion fit to
$\sqrt{\Delta_{min}^2+A^2\,(\ell\!-\!\ell_{min})^2}$, where
$\ell_{min}\!=\!\frac{1}{2}$ for $\Delta_{\pm}$ and
$\frac{1}{2}\!\pm\!\delta k_{sp}$ for the 0.29~meV excitation
(phason), and $A_{\Delta_+,\Delta_-,ph}\!\!=\!41(5),
48(5),32(2)\,\mbox{meV}/\mbox{r.l.u.}$. The magnetic Bragg peaks
are at $\frac{1}{2}\!\pm\!\delta k_{sp}$ with $\delta
k_{sp}\!=\!0.015$. Dashed: dispersion assuming the $H\!=\!0$
spin-wave velocity, $A\!=\!96.3\,\mbox{meV}/\mbox{r.l.u.}$.}
\label{fig3}
\end{center}
\end{figure}

\begin{figure}[htb!]
\begin{center}
\epsfig{file=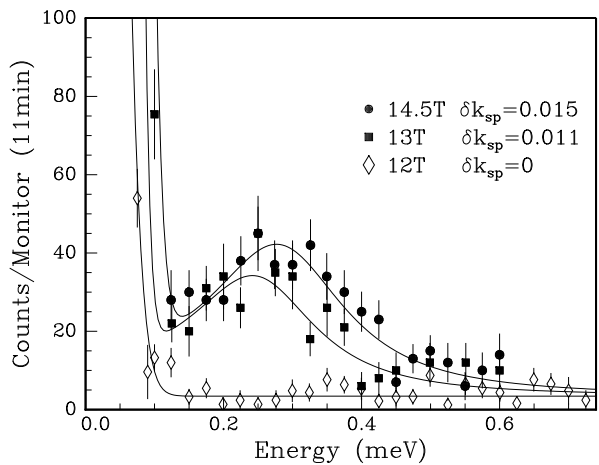,width=0.75\columnwidth,angle=90}  
\caption{Field dependence of the minimum energy of the phason,
energy scans at 12~T, 13~T and 14.5~T at the respective modulation
vector $(\frac12,1,\frac12+\delta k_{sp})$.} \label{fig4}
\end{center}
\end{figure}

A search was also made for excitations emanating from the magnetic
IC Bragg peaks at $(\frac12,1,\frac12\pm\delta k_{sp})$. This
revealed a mode at $\sim$0.29~meV (Fig.\ \ref{fig4}), at $T=2$~K
and $B=14.5$~T. The energy of this mode (at the respective IC
wave vector) changes only little between 13~T ($\delta
k_{sp}=0.011$) and 14.5~T ($\delta k_{sp}=0.015$). The dispersion
of this mode was studied at $B=14.5$~T and is plotted in the
bottom part of Fig.\ \ref{fig3}. Its spectral weight is
concentrated at the IC wave vector $(\frac12,1,\frac12\pm\delta
k_{sp})$, the minimum of its dispersion.

In interpreting the data the first thing to establish is whether
the three modes shown in Fig.\ \ref{fig3} are magnetic or nuclear
in origin. It is known from our earlier study that the satellites
at ($\frac12$,1,$\frac12\pm\delta k_{sp}$) are predominantly
magnetic (by a factor of $\sim$ 40) \cite{roe00}, and it follows
that the low-lying mode emanating from the satellites must also
have a predominantly magnetic character. For the higher-lying
modes it was observed that their intensity decreased at
equivalent $k$ but larger total wave vector transfer. They too
can therefore be ascribed to be mainly magnetic in origin. It
should be noted, however, that a structural contribution cannot
be excluded, and indeed a mixing of magnetic and elastic degrees
of freedom is expected in a spin-Peierls material.

Calculations of the excitations in the IC-phase are reported for
three cases, with dimerization parameters
$\epsilon=\frac{J_{i+1}-J_i}{J_{i+1}+J_i}=0.014$ \cite{poi97},
$\epsilon=0.14$ \cite{sch98d}, and $\epsilon=0.4$ \cite{yu99}.
Poilblanc \etal \cite{poi97} and Sch\"onfeld \etal \cite{sch98d}
also included frustrating next-nearest-neighbor intrachain
exchange $\alpha=\frac{2 J^{\text{nnn}}}{J_i+J_{i+1}}=0.24$ and
$\alpha=0.36$, while in the calculations of Yu \etal \cite{yu99}
only nearest-neighbor intrachain exchange was considered.
Irrespective of this large spread of the parameters $\epsilon$ and
$\alpha$ the spectral weight is concentrated at the commensurate
position $\ell=\frac12$ \cite{yu99,poi97}, where $\Delta_{\pm}$
have minimum excitation energy \cite{uhr98a}. This agrees with the
observed upper two modes at 1.3~meV and 1.7~meV (14.5~T, Fig.3),
which we therefore identify with the modes $\Delta_{\pm}$. The
longitudinal mode $\Delta_0$ should have its minimum at
$q_c=\frac12\pm\delta k_{sp}$ with
$\Delta_0^{min}\gtrsim\frac12(\Delta_+^{min}+\Delta_-^{min})$
\cite{yu99} and $g\mu_BH$ at $q_c=\frac12$ \cite{uhr98a}.
Therefore the 0.29~meV excitation cannot be identified with
$\Delta_0$. A weak longitudinal contribution at higher energies
could be hidden by the transverse excitations. However, the
weakness or absence of the longitudinal mode is probably related
to the significant interchain interaction in CuGeO$_3$.
Calculations \cite{yu99,poi97} which imply significant spectral
weight of the longitudinal mode do not include interchain
interaction, and it is known that longitudinal modes usually
become stronger with increasing one dimensionality of the exchange
\cite{ste86,af92b}.

While the field dependences of the modes $\Delta_{\pm}$ could in
principle be used to place further constraints on the values of
the coupling parameters, this turns out to be difficult in
practice. This is mainly due to two reasons, to the fact that the
values of $\epsilon$ used in the calculations \cite{yu99,sch98d}
lead to gaps much larger than reported here, and to the fact that
the magnetization range studied starts at larger values than the
maximum magnetization $m$=0.015 $\mu_B$ in our experiment. It is,
however, perhaps worth noting that the calculation with a finite
and large value of $\alpha$ \cite{sch98d} predicts that
$\Delta_{+}$ should increase with increasing field, while
$\Delta_{-}$ should be field independent, in qualitative
agreement with the data shown in Fig.\ \ref{fig1}. In this sense
our data are consistent with the large value of $\alpha$ needed
to explain various experimental facts
\cite{cas95,loo97,ara96,yok97b}, and with the notion that the
spin-Peierls transition in CuGeO$_3$ is driven by hard phonons
\cite{gro98,kub87,uhr98b}.

We now address the origin of the observed excitation at 0.29~meV.
Its field dependence is too small for an optic mode arising from a
folding back of the Brillouin zone. Such an optic mode should
shift with $\delta k_{sp}$ by about 0.1~meV from 13~T to 14.5~T. A
fit to a Lorentzian lineshape including the Bose factor for
$T=2$~K yields the finite energies shown in Fig.\ \ref{fig3}. The
same gap energy (2.3$\pm$0.2~cm$^{-1}\approx 0.29$~meV) was
observed in a Raman experiment \cite{loa99}, which mainly probes
the structural excitations, pointing to a magneto-structural
origin for this excitation. We attribute the mode at  0.29~meV to
phase oscillations of the soliton lattice, so-called phasons. The
IC long-range order in the high-field phase breaks the
quasi-continuous translation symmetry of the soliton lattice with
respect to simultaneous shifts of the soliton centers along the
chain axis. The Goldstone modes corresponding to this symmetry
breaking are phase oscillations of the magnetic and distortive
soliton structure. The corresponding eigenvectors carry magnetic
and elastic amplitudes. Their dispersion close to the IC
superlattice Bragg peak is expected to be given by the spin-wave
velocities of the dimerized phase, except close to the critical
field \cite{bha98}. Indeed the experimentally observed dispersion
along $\ell$ is of the same order of magnitude as that of the
transverse magnetic excitations (Fig.\ \ref{fig3}), and is flatter
by a factor of 3  than the zero-field spin-wave velocity.

If the mode at 0.29~meV is indeed a phason, it remains to be
explained why it has a gap. For a spin-Peierls system with soft
phonons, a first-order transition into the high-field phase could
proceed via a series of lock-in transitions, which implies a
finite phason gap \cite{lim83,bru78b}. No soft phonons have been
found in CuGeO$_3$, and a number of other possible mechanisms
need to be explored, including: impurities, discreteness of the
lattice, and $k$ dependent spin-phonon coupling. A pinning to
impurities would create a disordered soliton lattice, in
contradiction to our earlier structural study \cite{roe00}. The
discreteness of the lattice leads to a phason pinning energy of
the order 0.003$~$meV \cite{uhr99b}, which is much too small.
However, this estimate neglects interchain couplings, and does
not take into account the non-sinusoidal magnetic modulation
\cite{roe00}, both of which may enhance the phason gap in
CuGeO$_3$. In the absence of soft-phonons, lock-in phenomena in
CuGeO$_3$ could be produced by a $k$ dependent spin-phonon
coupling. Sch\"onfeld \etal \cite{sch98d} introduce a
$k$-dependent effective elastic constant $K(k)=\sum_{\lambda}
\Omega_{\lambda}(k)/g^2_{\lambda}(k)$, where $\Omega_{\lambda}(k)$
are the dimerization phonons and $g_{\lambda}(k)$ the respective
coupling constants. The relevant phonons in CuGeO$_3$ do not
display a softening in the dispersion at the commensurate wave
vector $k_{sp}$ \cite{bra98a}, but the spin-phonon coupling
increases towards $k_{sp}$ ($K(k_{sp})\approx 4K(0)$
\cite{wer99}). This preference for a commensurate structure
enhances the influence of the discreteness of the lattice, and
may lead to a finite phason energy and a commensurate wave vector,
instead of gapless phasons and a truely IC modulation.

The hysteresis observed in ESR experiments in the whole IC-phase
\cite{pal96a} supports the picture of a lock-in of the soliton
centers to the discrete lattice. Nevertheless, in the past,
several experimental results were taken as evidence for gapless
phasons: i) an increase of the $T^3$-contribution of the specific
heat capacity in the IC-phase \cite{lor96}; ii) the absence of
discrete lines in the NMR-lineshape; iii) the reduced ratio
$m_s/m_u$ of the staggered and uniform magnetic modulation
amplitude ($m_s$, $m_u$) derived from NMR experiments. At closer
inspection i) could be explained by an altered elastic constant
in the IC-phase rather than by gapless phasons. Phasons with a
dispersion similar to magnons yield $T^{\alpha'}$-contributions
with $\alpha'=1$ or 2 corresponding to their 1D or at least 2D
density of states. ii) The NMR lineshape could equally well be
explained by an unresolved assembly of the expected $\sim$150
(discrete) lines within 0.5~T and hence is still consistent with
the presence of lock-in phenomena. iii) Zero-point fluctuations
of gapless phasons should lead to the same reduced $m_s/m_u$ in a
neutron experiment where no such reduction was observed. The
neutron result agrees within 20\% with field theory where phasons
are neglected. This shows that zero-point phason oscillations are
suppressed and hence indicates independently a finite energy gap
of the phasons.

In summary, we have observed the predicted transverse magnetic
excitations $\Delta_{\pm}$ in the IC phase of a spin-Peierls
material which correspond to the creation or destruction of a
domain-wall pair. The longitudinal magnetic excitation $\Delta_0$
was not found, probably due to the 3D interactions. The phason
mode of the IC modulated structure has also been observed, and is
gapped with a maximum spectral weight at the position of the IC
satellites, $(\frac12,1,\frac12\pm\delta k_{sp})$. We argue that
the finite energy of the phason mode is consistent with all of
the experimental data available to date. We hope that our data
will stimulate and guide the development of a microscopic model of
CuGeO$_3$ which will ultimately provide a consistent description
over the entire temperature-field phase diagram of this
unconventional spin-Peierls system.

We thank G.~Uhrig, R.A.~Cowley, W.J.L.~Buyers, and K.~Knorr for
helpful discussions, and TMR and BMBF for funding.

\end{document}